\documentclass{elsarticle}
\usepackage{enumerate}
\usepackage{amsmath,amsthm}
\usepackage{amssymb, latexsym}
\usepackage[mathscr]{euscript}
\usepackage{color}
\usepackage{array}
\usepackage{flafter}
\usepackage{layout}
\usepackage{graphicx}
\usepackage{epstopdf}
\usepackage[ansinew]{inputenc}
\usepackage{xfrac}
\usepackage{algorithm}
\theoremstyle{plain}
  \newtheorem{teo}{Theorem}
  
  \newtheorem{lema}{Lemma}
  \newtheorem{prop}{Proposition}
  
\theoremstyle{definition}
  \newtheorem{defi}{Definition}
  
\theoremstyle{remark}

  \makeatletter
\def\ps@pprintTitle{%
   \let\@oddhead\@empty
   \let\@evenhead\@empty
   \def\@oddfoot{\reset@font\hfil\thepage\hfil}
   \let\@evenfoot\@oddfoot
}
\makeatother

\title{Mode hunting through active information}

\author[bst]{Daniel Andr\'es D\'{\i}az--Pach\'on\corref{cor1}}\ead{Ddiaz3@miami.edu}

\author[ie]{Juan Pablo S\'aenz}\ead{j.saenz4@umiami.edu}

\author[bst]{J. Sunil Rao\corref{cor2}}\ead{JRao@biostat.med.miami.edu}

\author[cwr]{Jean-Eudes Dazard}\ead{jean-eudes.dazard@case.edu}

\cortext[cor1]{Corresponding author}
\cortext[cor2]{Principal corresponding author}
\address[bst]{Division of Biostatistics - University of Miami, Don Soffer Clinical Research Center, 1120 NW 14th St, Miami FL, 33136}
\address[ie]{Department of Industrial Engineering - University of Miami, McArthur Engineering Building, 1251 Memorial Drive, Coral Gables FL, 33146}
\address[cwr]{Center for Proteomics and Bioinformatics, Case Western Reserve University, 10900 Euclid Av. Cleveland, OH 44106}

\begin{document}

\begin{abstract}
We propose a new method to find modes based on active information. We develop an algorithm that, when applied to the whole space, will say whether there are any modes present \textit{and} where they are; this algorithm will reduce the dimensionality without resorting to Principal Components; and more importantly, population-wise, will not detect modes when they are not present.
\end{abstract}

\maketitle

\section{Introduction}

When Newton and Leibniz developed Calculus independently during the seventeenth century AD \cite{Leibniz1684, Newton1736}, one of their main goals was to find a way to determine the position of maxima (or minima) in an Euclidean space. Of course, their interest in constructing a mathematical theory to discover maxima shows the already existent need to actually recognize these critical points ---a fact well illustrated by the usage Newton made of his own ``fluxions and infinite series'' when a few decades later he proposed physical mechanics to explain gravitation in his \textit{Principia} \cite{Newton1687}, arguably the most important scientific book of all times (see, e.g., \cite{Berlinski2002}).

Since then, the vast majority of the scientific endeavor has been related in one way or another to finding these maxima. In statistics, modes (and generalizations to regions with high accumulation of points) have been always central to the theory. In turn, modes have paved the way to introduce related concepts in data structures like bumps, components, clusters, or classes, among others. But chasing modes has proven to be extremely difficult, and large amounts of statistical research are devoted to improve already existing methods or developing new ones that allow a more efficient way to tackle the problem. Due to the advent of big data, bump hunting has thus acquired more and more importance. In this paper, we are primarily concerned with the particular case of \textit{mode} hunting (so from here on, every time that we are talking about \textit{bumps}, we refer in particular to \textit{modes}, a language convenience to avoid redundancies.)  

Parametrical methods work only relatively well in low dimensions. The way they work is rather straightforward: we find the distribution of a sample, say through a goodness of fit test, so then we know the position of its modes. For instance, if we find that our sample is distributed like a $\mathcal N(\mu, \sigma^2)$, we will know that there is a bump at $\mu$. Thus, it is not difficult to see how big data and large dimensionality will render useless most of these approaches. For this reason, non-parametrical approaches have become the default, even though these are also affected by the curse of dimensionality, among other problems in large dimensions. Our proposed strategy takes elements from both: like a goodness of fit test, it will compare the empirical distribution to a uniform one, if there are any differences between the two distributions, there is a mode. Like non-parametrical methods it will go for the particular modes to chase them. Therefore, the algorithm we present here will say whether there are any bumps present \textit{and} where they are. 

When $p \gg 0$, it is usually the case that parsimony is important in the sense that most of the interest is reduced to a few $p'$ variables such that $0 < p' \ll p$. The usual strategy used to reduce the dimensionality of the original information is to apply principal components (see. e.g. \cite{DiazDazardRao2016}), but this kind of reduction presents several problems in that the response might be associated to the zeroed dimensions \cite{HadiLing1998, Joliffe1982}. The algorithm we propose here reduces the dimensionality with no resource to principal components. This is the goal of Section 3.

Maybe the best known bump hunting algorithm is the Patient Rule Induction Method (PRIM) \cite{FriedmanFisher1999}. PRIM is a greedy algorithm made of three stages: peeling, pasting, and covering. The most important to understand here is peeling: The data lives in a $p$-dimensional box $B$ and, among a set $\mathcal B$ of elegible sub-boxes, we are going to remove the sub-box $b^*$ that maximizes the probability of $B-b$, for $b \in \mathcal B$. The boxes in $\mathcal B$ are the ones at the ``extreme'' of each dimension $j$; i.e., in each dimension we consider $b_{j^-} = \{\textbf x : x_j < x_{j(\alpha)} \}$ and $b_{j^+} = \{\textbf x : x_j > x_{j(1- \alpha)} \}$, where $x_{j(\alpha)}$ and $x_{j(1- \alpha)}$ are, respectively, the $\alpha$ quantiles and $(1-\alpha)$ quantiles. Then $\mathcal B$ is made of $2p$ boxes. Once we have $B-b^*$ we iterate the same procedure but now replacing $B$ by $B-b^*$ and choosing the subboxes among the ones defined by the $\alpha$ and $(1-\alpha)$ quantiles in $B-b^*$. We iterate the process until the peeled box $B^*$ reaches some probability threshold $\beta$. At the end, according to PRIM, there is a bump in $B^*$.

However, PRIM has at least two limitations: it becomes computationally expensive in high dimensions and, as shown by Polonik and Wang, it has problems in two or more dimensions because ``PRIM might not be able to resolve two distinct modes'' \cite{PolonikWang2010}. The second problem persists even when the tuning meta-parameters of the algorithm are chosen carefully. This means that in the end PRIM identifies as modes some things that are not modes. By contrast, the algorithm we propose here finds modes whenever they are present, up to the tuning of a parameter we call $b$, and we can be sure that it won't identify as modes anything that is not a mode.

In Section 3 we introduce our algorithm formally, in Section 4 we prove it is consistent. Section 5 presents some examples that illustrate the reach of the algorithm proposed. Section 6 uses the fact that low probability makes for large information, so that the curse of dimensionality in probabilistic terms can be seen also as a bless in terms of information. As for this Introduction, we now proceed to explain the basic facts of what we call \textit{active information} and its central role in mode hunting.

\section{Active information}

Conservation of Information (CoI) is a concept that has been around since the nineteenth century when Lady Lovelace for the first time alluded to it. Bringsjord et al. paraphrase her like this \cite{BringsjordEtAl2001}: 

\begin{quote}
	\textit{Computers can't create anything. For creation requires, minimally, originating something. But computers originate nothing; they merely 		do that which we order them, via programs, to do.}
\end{quote}

Almost two centuries after Lady Lovelace, many statements on CoI have been posed but they have only strengthen the case on information being conserved. The more definite affirmation came right at the end of the twentieth century, when Wolpert and Macready proved their so-called No Free Lunch Theorems (NFLT), in which they showed that, in the absence of additional information, no computer search is better on average than a blind search \cite{WolpertMacReady1995, WolpertMacReady1997}. But in real life some searchs do better than others. Why? Because the programmer is adding the relevant information given what he knows about the search. In Wolpert and MacReady's words, a search can be improved only by ``incorporating problem-specific knowledge into the behavior of the [optimization or search] algorithm.'' 

Active information (AI), following on the reasoning subjacent to the NFLT, has been introduced in order to measure how much information is being added by the programmer in a computer search \cite{DembskiMarks2009b}. In what remains of this section, we will introduce it in terms of mode hunting.

Assume a sample space $\Omega = \{1,\ldots, N\}$, with $N\in \mathbb N$. Since, by its very definition, the uniform distribution over $\Omega$ is the only one with total absence of bumps, we know that any nonuniform distribution over $\Omega$ \emph{do} forms bumps. This simple idea can be taken as a proof for the existence of bumps ---use any of the goodness-of-fit tests in the market and compare against uniformity. If the difference is significative, reject the hypothesis of uniformity, which is equivalent to accept that the underlying distribution has bumps.  This is useful by itself, but would be incomplete if it does not say \emph{where }the bumps are. A way to detect their location follows: 

Denoting $\textbf U$ as the uniform distribution, let's define 
\begin{align*}
	I_{\Omega} = -\log \textbf U(\omega)
\end{align*}
as the \textit{endogenous information} of the event $\omega$. Denoting now as $\textbf S$ the empirical distribution coming from the sample, define

\begin{align*}
	I_S = -\log \textbf S(\omega)
\end{align*}
as the \textit{exogenous information} of the event $\omega$. Finally, we can define the \textit{active information} of the event $\omega$ as \cite{DembskiMarks2009b}:
\begin{align*}
	I_+ := I_{\Omega} - I_S = \log\frac{\textbf S(\omega)}{\textbf U(\omega)}.
\end{align*}

With this apparatus, we propose to calculate the active information of each singleton event in $\Omega$. Modes, whenever they are present, would be located at events such that their active information is positive. In fact, the bigger the active information, the bigger the mode is going to be.

%

\section{Active Information Mode Hunting (AIMH) Algorithm}

We mention at the beginning of the Introduction that standard parametrical approaches do not use to do well when big data and/or high dimensions are involved, so non-parametrical methods are preferred when we are chasing bumps. This is one of the main reasons to introduce algorithmic-type methods like PRIM \cite{FriedmanFisher1999} or Local Sparse Bump Hunting \cite{DazardRao2010}. In the same spirit, we present here an algorithm we have dubbed Active Information Mode Hunting (AIMH) in which we mix the parametrical idea of a goodness-of-fit test to compare the data against a normal, with the non-parametrical advantage that is offered by an algorithmic approach.

Let's assume a $p$-dimensional rectangular space of finite Lebesgue measure $\textbf I_p = I_1 \times \cdots \times I_p$. Then the r.v. with maximum entropy is $\textbf U(\textbf I_p)$; i.e., a $p$-dimensional uniform distribution whose parameter is the Lebesgue measure of $\textbf I_p$. Consider $\textbf D :=\{1,\ldots, p\}$, and take a subset $\textbf D' :=\{i_1,\ldots, i_{p'}\}$, with $p' <p$, then $U_{i_1,\ldots, i_{p'}}$ is a projection of the uniform in $p$ dimensions in $p'$ dimensions. With this, we can extend the idea of the previous Section to find the $p$-dimensional modes as well as the modes in the projections through Algorithm \ref{AIMH} in page \pageref{AIMH}.

\begin{algorithm}[!ht]
\caption{Active Information Mode Hunting (AIMH)}\label{AIMH}
\begin{itemize}
	\item For $i$ from 1 to $p$: 
		\begin{itemize}
			\item Partitionate the interval $I_i$ into $r_i$ subintervals $I_{ij}$, with $j = 1,\dots, r_i$.
			\end{itemize}
	
	\item For $i$ from 1 to $p$:
		\begin{itemize}
			\item For $j$ from 1 to $r_i$:
				\begin{itemize}
					\item Calculate the AI:
						\begin{align*}
							I_+(I_{ij}) = \log \frac{P_s[X_s \in I_{ij}]}{P[U_i \in I_{ij}]},
						\end{align*} 
						where $X_s$ is the r.v.\ of the empirical distribution.
				\end{itemize}
			\item Define $\mathcal C_i = \{I_{ij} : I_+(I_{ij}) > b_1;\, j=1,\dots,r\}$, for a pre-specified $b_1 \in \mathbb R^+$.  
		\end{itemize}
	
	\item Define $\mathcal C^{(1)} := \{\mathcal C_i : i= 1,\ldots p\}$.
	\item For $k$ from 2 to $p$:
		\begin{itemize}
			\item Take every $k$-dimensional hyper-rectangle $I_{i_1j_1} \times \cdots \times I_{i_{k-1}j_{k-1}} \times I_{i_kj_k} \in \mathcal 					C^{(k-1)} \times \mathcal C^{(1)}$ and calculate its AI:
				\begin{align*}
					I_+\left(I_{i_1j_1} \times \cdots \times I_{i_kj_k}\right) = 
						\log \frac{P_s[X_s \in I_{i_1j_1} \times \cdots \times I_{i_kj_k}]}{P[U_{i_1, \dots, i_k} \in I_{i_1j_1} \times \cdots \times I_{i_kj_k}]}.
				\end{align*}
			\item Define
				\begin{align*}
					\mathcal C_{i_1\ldots i_k} = \big \{& I_{i_1j_1} \times  \cdots \times I_{i_{k-1}j_{k-1}} \times I_{i_kj_k} \in \mathcal C^{(k-1)} 							\times \mathcal C^{(1)} :\\ 
						& I_+\left(I_{i_1j_1} \times \cdots \times I_{i_kj_k}\right) > b_k;\, i_k \notin \{i_1,\ldots, i_{k-1}\} \text{ and }\\  
						& j_\ell = 1,\ldots, r_\ell, \text { for } \ell = 1,\ldots, k\big\},
				\end{align*}
				for a pre-specified $b_k \in \mathbb R^+$.
				\item Define $\mathcal C^{(k)} = \bigcup_{\mathcal I} \mathcal C_{i_1\ldots i_k}$, where $\mathcal I$ comprises the 								$\binom{p}{k}$ cases of different $\mathcal C_{i_1\ldots, i_k}$.
				\item If $\mathcal C^{(k)} = \emptyset$, halt.
		\end{itemize}
\end{itemize}
\end{algorithm}

Then $\mathcal C^{(i)}$ is the collection of bumps in $i$ dimensions found with this method, where $i=1\ldots, p$.

The algorithm is better understood informally in plain English:

\begin{enumerate}
	\item Partitionate every single dimension into subintervals.
	\item Calculate the AI of each subinterval in each dimension.
	\item In each dimension, keep those subintervals whose AI is bigger than some threshold $b_1$ (In fact, this can be generalized a little further 			considering different thresholds in every dimension $b_{1i}$, for $i$ from 1 to $p$.)
	\item For $i \neq j$, take all the subintervals in dimensions $i$ and $j$ whose AI was above the threshold $b_1$ (or $b_{1i}$ and $b_{1j}$ in 			the generalization) and construct bi-dimensional boxes.
	\item Define a threshold $b_2$ (or $b_{2ij}$) and collect the bi-dimensional boxes in dimensions $i$ and $j$ whose AI is above the threshold.
	\item Construct three-dimensional cubes with the boxes whose AI was above the threshold $b_2$ and the subintervals whose AI was above 			the threshold $b_1$.
	\item Collect the cubes whose AI is above a threshold $b_3$.
	\item Repeat recurrently the procedure adding the significative unidimensional subintervals to the significative hyper-cubes, to form hyper-			cubes with one additional dimension, either until exhaustion of dimensions or until there is a particular number of dimensions in which no 			hyper-cube has AI higher than the specified threshold.
\end{enumerate}

Several remarks are in order.

First, in some cases, it is not necessary to measure the AI against the uniform distribution, since the background distribution is	already given. In these cases, all that is needed is simply to replace $\textbf U$ by the relevant probability, say $\textbf Q$. In this case, the algorithm presented also works replacing in each dimension the marginal uniform by the corresponding marginal distribution.

Second, we can also consider more general $\sigma$-finite spaces, replacing the endogenous 	distribution by a distribution of interest. If the due knowledge about the moments is given, this endogenous distribution can take the form of a maximum entropy distribution (see, e.g., \cite{Diazmarks2017}). In reality, however, this is only of theoretical interest, since in applications all the spaces will be effectively finite, therefore uniformity will do as background in more computational implementations.

Third, the algorithm depends on the choice of the metaparameters $r_i$ (the number of intervals in the partition per dimension) and $b_i$, for 		$i = 1,\ldots, p$. Under reasonable assumptions, it is expected of Algorithm \ref{AIMH} to diminish the size of the space considered at a fast speed. If it is reasonable to set every dimension to have the same amount of subintervals (i.e., $r = r_1 = \cdots = r_p$), then it also makes sense to define $b_{k+1} := b_kr^{k+1}$, so that actually we only need to define $b_1$. From now on, we use this assumption.

\section{Consistency}

In this section we prove that Algorithm \ref{AIMH} in Section 3 estimates consistently the modes of the population distribution function. Without loss of generality, we assume that the underlying space is $[0,1]^p$, and, as a simplifying assumption, we partition every dimension in the same amount of subintervals, i.e., $r_1 = \cdots = r_p = r$.

Let's define first a more general version of AI with respect to a different underlying distribution:

\begin{defi}[\textbf{AI under $X$ with respect to $Y$}]
	Let $X$ and $Y$ be continuous r.v.'s over the same sample space $\Omega$, and let $A$ in $\Omega$ be a Borel set such that $P_Y(A) > 0$, then
	\begin{align*}
		I_+^{X | Y}(A) = \log \frac{P_X(A)}{P_Y(A)}.
	\end{align*}
\end{defi}

With this definition we prove the following result, from which consistence is derived:

\begin{teo}\label{ConsisTheo}
	Let $F_n$ be an empirical distribution function coming from the continuous distribution $F$ over the sample space $[0,1]$. The AI under 			$F_n$ with respect to $U$ consistently estimates the AI under $F$ with respect to $U$:
	\begin{align}\label{Consistency}
		I_+^{F_n | U}(A) \xrightarrow{n \rightarrow \infty} I_+^{F | U}(A), \text{\ \ \  a.s. }
	\end{align}
\end{teo}

In order to prove Theorem \ref{ConsisTheo}, for a vector $\textbf x = (x_1, \ldots, x_p) \subset [0,1]^p$, we define first $[0, \textbf x]^p$ as the $p$-dimensional box $[0, x_1] \times \cdots \times [0, x_p]$. We have the following two lemmas:

\begin{lema}\label{AIConsist}
	Let $F_n$ be an empirical distribution function taken from a continuous distribution $F$ over the sample space $[0,1]$, then
	\begin{align*}
		\left|I_+^{F_n | F}([0,\textbf x]^p)\right| \xrightarrow{n\rightarrow \infty} 0, \text{\ \ \  a.s. }
	\end{align*} 
\end{lema}

\begin{proof}
	By the Glivenko-Cantelli Theorem, $F_n(x)$ converges to $F(x)$, with probability one, for every $\textbf x \in [0,1]^p$. Since $F_U(\textbf x) = x_1 \cdots x_p$, the quotient $F_n(\textbf x)/F_U(\textbf x)$ is always defined for $\textbf x \in (0,1]^p$. An application of the continuous mapping theorem completes the proof.
\end{proof}

AI satisfies this very convenient property:

\begin{lema}\label{Add}
Let $X, Y,$ and $Z$ be r.v. in $\Omega$. Let $A$ be a measurable set such that $F_Y(A)$ and $F_Z(A)$ are not zero. Then
	\begin{align*}
		I_+^{X | Z}(A) =  I_+^{X | Y}(A) + I_+^{Y | Z}(A). 
	\end{align*}
\end{lema}

\begin{proof}
	\begin{align*}
		I_+^{F | H}(A) =   \log \frac{P_X(A)}{P_Z(A)} = \log \frac{P_X(A)}{P_Y(A)} + \log \frac{P_Y(A)}{P_Z(A)}  = I_+^{F | G}(A) + I_+^{G | H}(A).
	\end{align*}
\end{proof}
 
 Theorem \ref{ConsisTheo} follows now easily from the previous two lemmas:
 
\begin{proof}[\textbf{Proof of Theorem \ref{ConsisTheo}}]
	Let $X_n$ be the r.v.\ associated to $F_n$, $X$ the r.v. associated to the limiting distribution $F$, and $U$ a uniform in $[0,1]$. By Lemma 				\ref{Add}, we obtain for $A = [0, \textbf x]^p$ that
	\begin{align*}
		I_+^{F_n | U}(A)  = I_+^{F_n | F}(A) + I_+^{F | U}(A).
	\end{align*}
	By Lemma \ref{AIConsist}, as $n \rightarrow \infty$, we obtain then that $I_+^{F_n | U}(A) \rightarrow I_+^{F | U}(A)$. 
	Since the class $\{ [0, \textbf x]^p : \textbf x \in [0,1]^p\}$ is a $\pi$-system that generates the Borel $\sigma$-algebra, the monotone class 		Theorem (Dynkin's Theorem) generalizes the result to every Borel set $A$ such that $F(A) \neq 0 \neq |A|$.
\end{proof}

\section{Examples}

In this section we examine three examples. The first illustrates in one dimension how our algorithm works, the last two explain in two dimension the workings of the algorithm. The third in particular is very interesting, since it shows its superiority to PRIM in that it finds bumps that PRIM is not able to detect. 

A nice feature of Algorithm AIMH is that the AI of each box can be calculated exactly due to the fact that we are using the probability induced by the empirical distribution and the uniform one. Or in the following examples, we can limit ourselves to compare the AI in the relevant boxes based on the probabilities of the regions under the parametrical endogenous and exogenous distributions. Therefore, no simulations are needed.

\subsection{Normal of low variance $N(0,10)$}

In this example we examine the active information in an interval centered around the mean.

Let $X\sim N(0,10)$ restricted to $\Omega = [-8.5, 8.5]$. Call $A = [-.5, .5]$. Thus, $P[X \in A] = 0.0659$. From here we can calculate the active information of the event $A$ as follows:
\begin{align*}
	I_+(A) &\approx \log_2 1.1203\\
		&\approx 0.1638.
\end{align*}

Then, for $r = 1/17$ and any $b > .163$ (a very small bound!), our algorithm does conclude that there is no bump in $A$. When we realize that $P[X \in A]$, and $P[U \in A]$ are pretty close (0.0659 and 0.0526, respectively), it does not look that surprising. However, if we are sure that there is a mode (like in this case), we can always lower our bound. In this case $b= 0.16$ will find a bump for the interval $A$.

This example shows how the process works in one dimension: of all the possible subintervals we consider only the ones in which the 			active information has more bits than $b_1$.

\subsection{Bivariate normal $N((0,0), \textbf I)$}

Consider $\textbf Z \sim N((0,0), \textbf I)$, with marginals $X$ and $Y$. Call $A = [-.5, .5]$. Then
\begin{align*}
	P(X \in A) = P(Y \in A) \approx 0.3829.
\end{align*}

Considering again the interval $\Omega = [-8.5, 8.5]$ in each dimension, we have that
\begin{align*}
	I_+(A) &= \log_2(.3829 \times 17)\\
		&= \log_2(6.5093)\\
		&= 2.7.
\end{align*}

Therefore, if our $b_1 > 2.5$, then the interval $A$ in each dimension is collected. Then we make the Cartesian product $A\times A$ and calculate its AI:
\begin{align*}
	I_+(A \times A) &\approx \log_2 (0.1444 \times 289)\\
			&\approx \log_2(41.73)\\
			&\approx 5.38.
\end{align*}

This example illustrates how to apply the algorithm when $p = 2$. It reveals the nice feature of Algorithm \ref{AIMH} that it does not even need to	consider a whole dimension, but only the portion in each dimension that if finds relevant. That is why we only consider the subintervals centered around the mean to construct the boxes in dimension 2, because only those had information bigger than 2.5. 

The example also illustrates the importance of considering an algorithmic method, since choosing first all subintervals in every dimension (which in this example amounts to one subinterval in each dimension) allows us to go in a second step to make boxes to look at the AI in them.

\subsection{Bivariate mixture of normals}

Consider two bivariate random vectors $\textbf X\sim N\left((-3, 0),  \frac{1}{4}\textbf I\right)$ and $\textbf Y\sim N\left((3,0), \frac{1}{4}\textbf I\right)$ with distributions $f_\textbf X$ and $f_\textbf Y$, respectively. Consider the new random vector $\textbf Z$ with density
\begin{align*}
	f_\textbf Z = \frac{1}{2}f_\textbf X + \frac{1}{2}f_\textbf Y
\end{align*}

Thus each the marginal density of $Z_1$ is given by $f_{Z_1} = \frac{1}{2}f_{X_1} + \frac{1}{2} f_{Y_1}$, and the marginal $Z_2 \sim N(0,1/4)$. We also consider the space $\Omega = [-8.5, 8.5]$ in the first axis, and $[-4.5, 4.5]$ in the second one. 

Then partitioning each dimension in intervals of length one (centered around every integer), and taking $b_1 = 2.6$, we obtain:
\begin{align*}
	P(Z_1 \in [-3.5, -2.5]) &= 0.5(0.6827) + 0.5P(Y_1 \in [-3.5, -2.5])\\
				& \approx 0.5(0..6827) + \varepsilon\\
				&\approx 0.3413 + \varepsilon\\
	P(Z_1 \in [2.5, 3.5]) &\approx 0.3413 + \varepsilon.
\end{align*}

For $P(Z_2 \in [-.5, .5]) \approx 0.6827$.

The active information of these intervals is as follows:
\begin{align*}
	I_+^{Z_1}([2.5, 3.5]) &\approx \log_2 (0.3413 \times 19)\\
					&\approx \log_2 6.48\\
					&\approx 2.69.\\
	I_+^{Z_2}([-0.5, 0.5]) &\approx \log_2(0.6827 \times 9)	\\
					&\approx \log_2 6.14	\\
					&\approx 2.61.	
\end{align*}

No other intervals have active information bigger than 2.6. Therefore in $Z_1$ we take the intervals $[-3.5, 2.5]$ and $[2.5, 3.5]$; in $Z_2$ we take the interval $[-0.5, 0.5]$.

Having done this, we now consider the active information inside the Cartesian products $ A = [-3.5, -2.5] \times [-0.5, 0.5]$ and $B = [2.5, 3.5] \times [-0.5, 0.5]$.
\begin{align*}
 I_+(A)  = I_+(B) &\approx \log_2 (0.4577 \times 171)\\
 	 &\approx 6.29.
\end{align*}

This example is interesting and important. Polonik and Wang illustrated through some simulations the incapacity of PRIM in two or more dimensions to differentiate the two picks of a bimodal symmetric mixture of two normals; PRIM finds a single region containing the two bumps and the valley between them (see Figure 3 in \cite{PolonikWang2010}). Our example has the same characteristics as the distribution considered in \cite{PolonikWang2010}, and shows that, given an adequate tuning of the parameters, active information is able to differentiate the two picks through two different boxes. In fact, in general, active information is able to find as many modes as there are present, provided the right choice of the parameters.

As in the previous example, even with only two dimensions, this one again illustrates the importance of going algorithmically increasing the dimension of the boxes that are relevant in terms of AI.

\section{The bless of dimensionalty}

As it is well known, as long as the number of dimensions is increasing, and the number of observations is fixed, the curse of dimensionality will make it difficult to find uniform distributions (see, e.g., \cite{HastieTibshiraniFriedman2009}, pp. 22--27). The easy way to see this is to consider a discrete uniform r.v. with parameter $N$ in a sample space $\Omega$. Then, the random vector whose components are iid uniform r.v. living each in $\Omega$ lives itself in the Cartesian product $\Omega \times \cdots \times \Omega$ ($p$ times), so that each point has probability $N^{-p}$. Thus most, if not all, learning methods ---and this is not restricted to bump hunting--- will fail to detect uniform distributions in high dimensions, unless the sample size is growing exponentially with the number of dimensions.

On the other hand, this exact problem serves well in terms of information ---lower probability always makes for logarithmically higher information. We can state formally the previous assertion in terms of the following Proposition:

\begin{prop}
	When we are considering the active information of a random vector whose components are iid uniform random variables distributed uniformly 		in a sample space $\Omega$, the active information of the random vector becomes a linear function of $p$ with slope $-I_\Omega$ and 			intercept $-I_S$.
\end{prop}

\begin{proof}
	\begin{align*}
		I_+ &= I_{\Omega \times \cdots \times \Omega} - I_S\\
			&= -\log N^{-p} - I_S\\
			&= -p I_\Omega - I_S. 
	\end{align*}
\end{proof}

\section{Discussion}

We have proposed the AIMH algorithm, which consistently finds bumps, whenever they are present, and which informs of the lack of bumps when these are not there. Additionally, our proposed algorithm works well in high dimensional spaces as it reduces the dimensionality without resorting to Principal Components. The AIMH algorithm takes elements from both parametric and non-parametric methods to determine the presence and locations of modes. Furthermore, the AIMH algorithm has distinct advantages over the PRIM algorithm as it is not as computationally expensive in high dimensions, and will always find the modes given that the parameters $r_i$, and $b_i$ are tuned accurately.

The tuning of $r_i$, and $b_i$ when data is involved, exceeds the scope of this research and will be subject to further study. In applied cases, these parameters need to be determined with the existing data via cross-validation or by the researcher's knowhow using Bayesian inference.

\bibliographystyle{plain}
\bibliography{MyBibliography}

\begin{thebibliography}{10}

\bibitem{Berlinski2002}
David Berlinski.
\newblock {\em Newton's Gift: How Sir Isaac Newton Unlocked The System of the
  World}.
\newblock Free Press, 2002.

\bibitem{BringsjordEtAl2001}
S.~Bringsjord, P.~Bello, and D.~Ferrucci.
\newblock Creativity, the turing test, and the (better) lovelace test.
\newblock {\em Minds and Machines}, 11(1):3--27, 2001.

\bibitem{DazardRao2010}
J-E. Dazard and J.~S. Rao.
\newblock Local {S}parse {B}ump {H}unting.
\newblock {\em J. Comput. Graph. Stat.}, 19(4):900--929, 2010.

\bibitem{DembskiMarks2009b}
W.~A. Dembski and R.~J. Marks.
\newblock Conservation of {I}nformation in {S}earch: {M}easuring the {C}ost of
  {S}uccess.
\newblock {\em IEEE Transactions on Systems, Man and Cybernetics A, Systems \&
  Humans}, 5(5):1051--1061, September 2009.

\bibitem{DiazDazardRao2016}
D.~A. D\'{\i}az-Pach\'on, J-E. Dazard, and J.~S. Rao.
\newblock Unsupervised bump hunting using principal components.
\newblock In S.E. Ahmed, editor, {\em Big and Complex Data Analysis:
  Methodologies and Applications}. Springer, 2016.

\bibitem{Diazmarks2017}
D.~A. D\'{\i}az-Pach\'on and R.~J. Marks.
\newblock Maximum entropy and active information.
\newblock {\em Submitted}, 2017.

\bibitem{FriedmanFisher1999}
J.H. Friedman and N.I. Fisher.
\newblock Bump hunting in high-dimensional data.
\newblock {\em Stat. Comput.}, 9:123--143, 1999.

\bibitem{HadiLing1998}
S.~Hadi and R.F. Ling.
\newblock Some cautionary notes on the use of principal components regression.
\newblock {\em Am. Stat.}, 52(1):15--19, 1998.

\bibitem{HastieTibshiraniFriedman2009}
T.~Hastie, R.~Tibshirani, and J.~Friedman.
\newblock {\em The Elements of Statistical Learning: Data Mining, Inference,
  and Prediction}.
\newblock Springer Science, New York, 2009.

\bibitem{Joliffe1982}
I.~Joliffe.
\newblock A note on the use of principal components in regression.
\newblock {\em J. R. Stat. Soc. Ser. C. Appl. Stat.}, 31(3):300--303, 1982.

\bibitem{Leibniz1684}
G.~Leibniz.
\newblock Nova methodus pro maximis et minimis.
\newblock {\em Acta Eroditorum}, (X):467--473, October 1684.

\bibitem{Newton1736}
I.~Newton.
\newblock {\em The Method of Fluxions and Infinite Series with its application
  to the Geometry of Curve-Lines}.
\newblock Henry Woodfall, 1736.

\bibitem{Newton1687}
Isaac Newton.
\newblock {\em Philosophi\ae Naturalis Principia Mathematica}.
\newblock 1687.

\bibitem{PolonikWang2010}
W.~Polonik and Z.~Wang.
\newblock {PRIM} analysis.
\newblock {\em J. Multivar. Anal.}, 101(3):525--540, 2010.

\bibitem{WolpertMacReady1995}
D~H Wolpert and W~G MacReady.
\newblock No free lunch theorems for search.
\newblock Technical Report SFI-TR-95-02-010, Santa Fe Institute, 1995.

\bibitem{WolpertMacReady1997}
D~H Wolpert and W~G MacReady.
\newblock No free lunch theorems for optimization.
\newblock {\em IEEE Transactions on Evolutionary Computation}, 1(1):67--82,
  1997.

\end{thebibliography}

\end{document}